\documentclass[usenatbib,]{mnras}

\makeatletter
\def\set@curr@file#1{%
  \begingroup
    \escapechar\m@ne
    \xdef\@curr@file{\expandafter\string\csname #1\endcsname}%
  \endgroup
}
\def\quote@name#1{"\quote@@name#1\@gobble""}
\def\quote@@name#1"{#1\quote@@name}
\def\unquote@name#1{\quote@@name#1\@gobble"}
\makeatother
\usepackage{graphicx}

\usepackage{amsmath}

\title[Slow Alfv\'enic solar wind origins]{The origin of slow Alfv\'enic solar wind at solar minimum}
\author[D. Stansby et al.]
{
	D. Stansby$^{1}$\thanks{E-mail: \href{d.stansby@ucl.ac.uk}{d.stansby@ucl.ac.uk}}\thanks{Now at Mullard Space Science Laboratory, University College London, Holmbury St. Mary, Surrey RH5 6NT, UK} ,
	L. Matteini$^{2, 1}$,
	T. S. Horbury$^{1}$,
	D. Perrone$^{1, 3}$,
	R. D'Amicis$^{4}$,
	L. Ber\v{c}i\v{c}$^{2,5}$
\\
$^{1}$Department of Physics, Imperial College London, London, SW7 2AZ, UK
\\
$^{2}$LESIA, Observatoire de Paris, Universit\'e PSL, CNRS, Sorbonne Universit\'e, \\~~Univ. Paris Diderot, Sorbonne Paris Cit\'e, 5 place Jules Janssen, 92195 Meudon, France
\\
$^{3}$ASI - Italian Space Agency, via del Politecnico snc, I-00133 Rome, Italy
\\
$^{4}$INAF - IAPS, Via Fosso del Cavaliere 100, Rome, Italy
\\
$^{5}$Physics and Astronomy Department, University of Florence, Via Giovanni Sansone 1, I-50019 Sesto Fiorentino, Italy
}
\pubyear{2019}

\begin{document}
\maketitle

\begin{abstract}
Although the origins of slow solar wind are unclear, there is increasing evidence that at least some of it is released in a steady state on over-expanded coronal hole magnetic field lines. This type of slow wind has similar properties to the fast solar wind, including strongly Alfv\'enic fluctuations. In this study a combination of proton, alpha particle, and electron measurements are used to investigate the kinetic properties of a single interval of slow Alfv\'enic wind at 0.35 AU. It is shown that this slow Alfv\'enic interval is characterised by high alpha particle abundances, pronounced alpha-proton differential streaming, strong proton beams, and large alpha to proton temperature ratios. These are all features observed consistently in the fast solar wind, adding evidence that at least some Alfv\'enic slow solar wind also originates in coronal holes. Observed differences between speed, mass flux, and electron temperature between slow Alfv\'enic and fast winds are explained by differing magnetic field geometry in the lower corona.
\end{abstract}


\section{Introduction}
The solar wind is an ionised plasma flowing at large speeds from the surface of the Sun to the edge of the Heliosphere. Although the density, speed, and temperature of the solar wind are all highly variable, it is possible to identify categories of solar wind with distinct properties. The most clearly defined category is that with the highest speeds, typically called the fast solar wind. Comparison between remote observations and in situ measurements show that fast solar wind originates in large coronal holes spanning over 60$^{\circ}$ in angular width \citep{Krieger1973, Nolte1976, Cranmer2009, Garton2018}.

In contrast the solar sources of wind with slow and intermediate speeds are varied and still not entirely clear \citep[e.g. see the review of][]{Abbo2016}. There are, however, multiple lines of evidence that a significant fraction of the slow solar wind also originates in coronal holes. One of the defining features of fast solar wind is a lack of variance in number density, velocity, and temperature, aside from pure Alfv\'enic fluctuations, which result in magnetic field and velocity variations that are either strongly correlated or anti-correlated \citep{Belcher1971, Bame1977}. A steady background state with superimposed Alfv\'en waves is also observed in situ in a large amount of the slower solar wind at all stages of the solar cycle \citep{Marsch1981, DAmicis2015a, Stansby2018c}, implying that close to the Sun it may also be heated and released into the heliosphere in a steady state manner on open field lines.

One significant difference between the regions of coronal holes that are thought to produce slow and fast winds is the magnetic field geometry in the corona. The amount of magnetic field expansion alters the location of the critical point where the plasma becomes supersonic \citep{Cranmer2005a}, which in turn alters the effects of heating processes. If the majority of heating happens below the critical point, the speed is not significantly affected, whereas significant heating above the critical point is expected to increase the speed \citep{Leer1980}. Because rapidly diverging magnetic fields have a higher critical point, more energy is deposited before the wind becomes supersonic, thus resulting in slower wind speeds \citep{Levine1977, Leer1980, Wang1991}.

These theoretical predictions agree well with an observed anti-correlation between solar wind speed at 1 AU and the amount of super-radial expansion the magnetic field undergoes close to the Sun \citep{Wang1990, Wang2006, Suzuki2006, Fujiki2015}. Statistically the smallest coronal holes, with the highest magnetic field expansions, produce wind with the slowest speeds \citep{Nolte1976, Hofmeister2018, Garton2018}, which has been verified on a case by case basis using magnetic field mappings from spacecraft to the solar surface \citep{Wang2009, Wang2017, Wang2019}. In addition, case studies at 1~AU show that the slow Alfv\'enic wind contains similar heavy ion composition \citep{DAmicis2018}, and similar alpha to proton abundance ratios \citep{Ohmi2004} as fast solar wind, reinforcing a probable similar solar origin. These previous studies have shown that the bulk properties of the slow Alfv\'enic wind are consistent with the theory that it originates in open field lines rooted in coronal holes.

If this theory is true, one would expect the processes occurring as the solar wind is heated and accelerated to be similar in both the fast and slow Alfv\'enic wind, and thus expect similar features to be found in the velocity distribution functions in both types of wind. The kinetic features of fast solar wind have been extensively characterised: it contains a proton beam population \citep{Feldman1973, Feldman1993}, large alpha particle to proton temperature ratios \citep{Marsch1982c, Stansby2019a}, magnetic field aligned proton-alpha particle streaming \citep{Marsch1982c, Neugebauer1996}, and large alpha to proton abundance ratios \citep{Aellig2001, Kasper2007a}. \cite{Marsch1982b} and \cite{DAmicis2018} have shown that at 1 AU both fast and slow Alfv\'enic winds tend to have isotropic proton distributions, but the other features have yet to be measured in slow Alfv\'enic wind.

In this paper we provide the first observations of these kinetic features in both slow and fast Alfv\'enic winds. Some of these features are destroyed by the time the solar wind has propagated to 1 AU, necessitating the use of data from the \emph{Helios} mission, which measured the solar wind from 0.3~--~1~AU. Three intervals are studied, initially identified by \cite{Stansby2018c} as a) typical fast solar wind, b) Alfv\'enic slow solar wind, and c) non-Alfv\'enic slow solar wind. Our comparison of both the bulk and kinetic features of fast and slow Alfv\'enic intervals shows that the slow Alfv\'enic period most likely originated from a small, over-expanded coronal hole.
\section{Data}
\label{sec:data}
The data used here were measured by the \emph{Helios} mission, which consisted of two spacecraft in heliocentric orbits between 0.3 and 1 AU \citep{PORSCHE1977}. The first perihelion pass of \emph{Helios 1} during solar minimum was used, from which three periods were chosen as representative examples of fast, slow Alfv\'enic and slow non-Alfv\'enic wind. The intervals are listed in table \ref{tab:intervals} and shown later in figure \ref{fig:timeseries}, and were chosen to contain continuous alpha particle measurements and to be long enough to build up a statistical characterisation of each interval.
\begin{table}
	\caption{Start and end times for selected intervals. All data are taken from \emph{Helios 1}.}
	\begin{tabular}{ccc}
		\hline
		Category		& Start time (UT)	& End time (UT)  \\ \hline
		Fast			& 1975/03/13 12:00& 1975/03/16 12:00			\\
		Slow Alfv\'enic	& 1975/03/23 00:00& 1975/03/24 16:00			\\
		Slow non-Alfv\'enic	& 1975/03/25 14:00	& 1975/03/26 00:00	\\
		\hline
	\end{tabular}
	\label{tab:intervals}
\end{table}

Particle data were measured by the E1 set of instruments, consisting of electrostatic analysers for both ions and electrons \citep{Schwenn1975}. Magnetic field data were measured by both the E2 and E3 fluxgate magnetometers \citep{Musmann1975, Scearce1975}. Where possible magnetic field data from E2 were used, with E3 data as a fallback. Although data gaps are frequent, the particle data is available at a maximum cadence of 40.5 seconds, and magnetic field measurements used here were averaged over the time taken to build up each individual particle distribution function.

Alpha particle parameters were calculated using the bi-Maxwellian fitting routine described in Appendix A of \cite{Stansby2019a}, with minor modifications to adapt the fitting to work well in both slow and fast solar winds\footnote{The updated fitting routine can be found at \url{https://github.com/dstansby/alphafit}, and the newly calculated parameters at \url{https://doi.org/10.5281/zenodo.3258337}}. Proton temperatures and velocities are bi-Maxwellian fits to the proton core from the dataset of \cite{Stansby2018h}\footnote{Available at \url{https://doi.org/10.5281/zenodo.1009506}.}.  Electron core parameters were obtained using the method presented in \cite{Bercic2019}, also calculated with a bi-Maxwellian fitting routine. For each species, the fitted parameters are number density ($n_{i}$), velocity ($\mathbf{v}_{i}$), and temperatures perpendicular and parallel to the local magnetic field ($T_{i\perp}, T_{i\parallel}$). The subscript $i$ is substituted as $p$ for protons, $\alpha$ for alpha particles, or $e$ for electrons. 

\begin{figure}
	\includegraphics[width=\columnwidth]{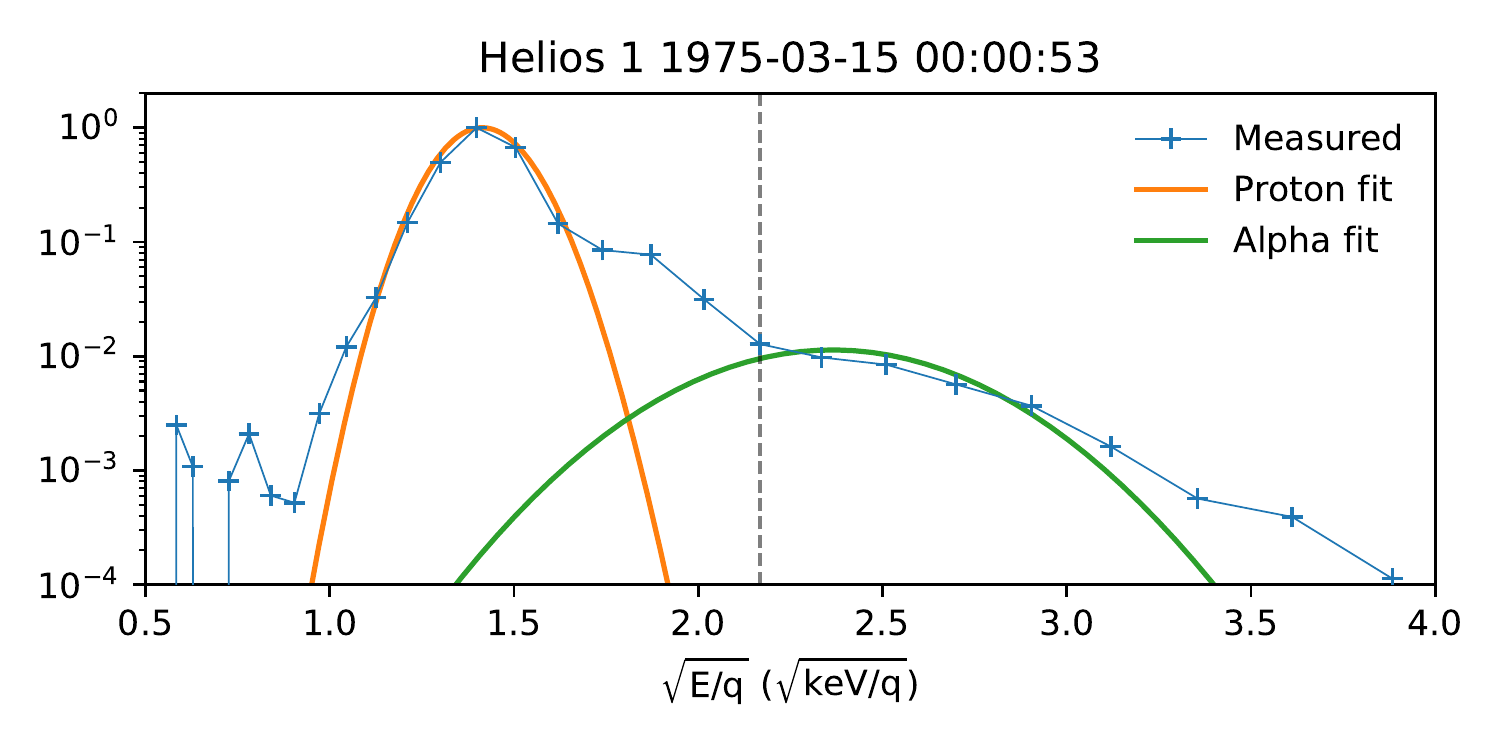}
	\includegraphics[width=0.495\columnwidth]{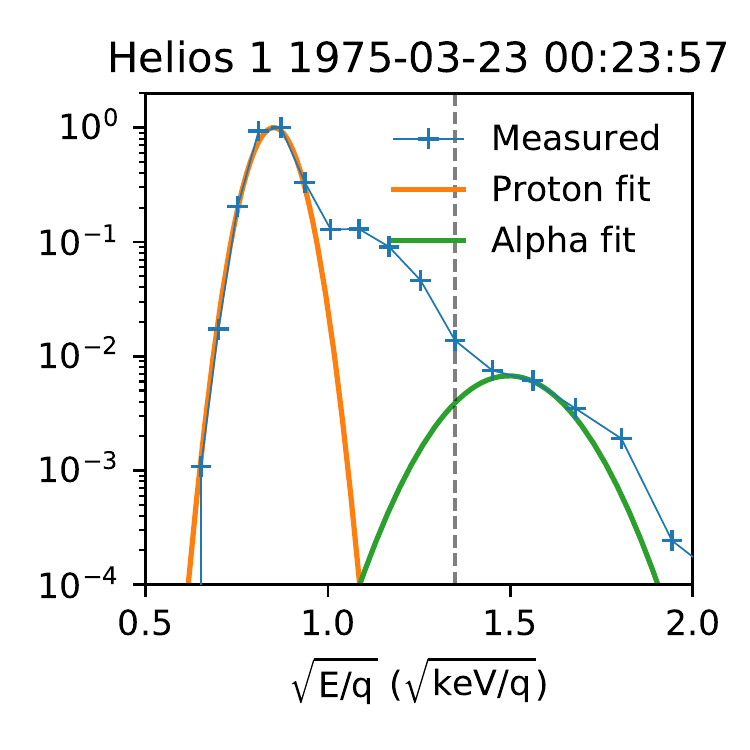}
	\includegraphics[width=0.495\columnwidth]{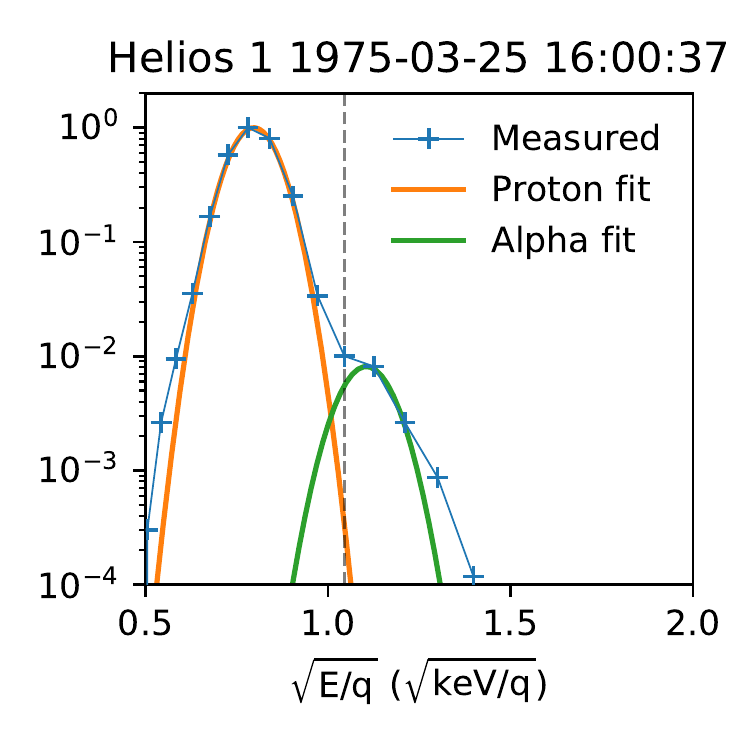}
	\caption{Example energy spectra (blue crosses) measured in each type of solar wind. Top panel shows fast wind, bottom left slow Alfv\'enic wind, and bottom right slow non-Alfv\'enic wind. The solid angle integrated bi-Maxwellian fits are shown for protons (orange line) and alpha particles (green line). The vertical grey line shows the dividing line between measurements dominated by protons (to the left) and those dominated by alpha particles (to the right). Note that the fits are performed in the full 3D velocity space, and shown here are 1D reductions of the distribution functions and fits.}
	\label{fig:dists}
\end{figure}

Example ion energy spectra for each interval, along with their corresponding proton and alpha particle fits, are shown in figure \ref{fig:dists}, demonstrating that the fitting works well in all types of wind. In both the proton and alpha parts of the distribution function a high energy tail or `beam' is present \citep[e.g.][]{Feldman1973, Marsch1982c, Marsch1982b} that by design is not captured by these fits to the core parts of the distribution function.

Energy per charge units (measured directly by the instrument) can be converted to velocity for a given particle species by
\begin{equation}
	\sqrt{\frac{E}{q}} = v \sqrt{\frac{m}{2q}}
\end{equation}
where $m$ and $q$ are the particle mass and electromagnetic charge respectively. Because of their increased charge to mass ratio, in these plots alpha particles appear at $\sqrt{E/q}$ values a factor of $\sqrt{2}$ higher than protons travelling at the same speed.

In figure \ref{fig:dists} some differences are already apparent between the three types of wind. The fast (top panel) and slow Alfv\'enic (bottom left panel) wind show a similar structure, with a proton core population, an additional proton beam population (not captured by the fits, but characterised later), and an alpha particle distribution which is wider and therefore significantly hotter than the proton distribution. In contrast the slow non-Alfv\'enic wind (bottom right panel) has no obvious proton beam, and a much thinner alpha particle distribution than the other types of wind.

In order to characterise the extra beam population of protons, the beam number density was calculated as the numerical integral
\begin{equation}
	n_{b} = \iiint \left [ f_{obs} \left (\mathbf{v} \right ) - f_{bimax} \left ( \mathbf{v} \right ) \right ] d\mathbf{v}
\end{equation}
where $f_{obs}$ is the observed velocity distribution function, $f_{bimax}$ is the fitted bi-Maxwellian core distribution function, and the integral is taken over the 3D velocity space with upper integration limit in $\left | \mathbf{v} \right |$ set to exclude the areas of $f_{obs}$ dominated by alpha particles. Unless otherwise stated, the total proton number density ($n_{p}$) is the sum of the proton core and beam number densities.

Aside from the aforementioned basic plasma parameters, some derived parameters are used later, which are defined as follows:
\begin{itemize}
	\item The Alfv\'en speed is
	\begin{equation}
		v_{A} = \frac{\left | \mathbf{B} \right |}{\sqrt{\mu_{0} \rho}}
	\end{equation}
	where $\rho = \sum_{i} n_{i} m_{i}$ is the total mass density of the plasma (including the proton beam).
	\item The proton beam fraction is
	\begin{equation}
		\frac{n_{b}}{n_{p}} = \frac{n_{b}}{n_{pc} + n_{b}}
	\end{equation}
	where $n_{pc}$ is the proton core number density calculated from bi-Maxwellian fits.
	
Finally, as a proxy for the Alfv\'enicity of the plasma, the cross helicity ($\sigma_{c}$) calculated from multiple measurements is used. This is defined as  \citep{Bruno2013a}
	\begin{equation}
	\sigma_{c} = 2 \frac{\left \langle \mathbf{v} \cdot \mathbf{b} \right \rangle}{\left \langle \left | \mathbf{v} \right |^{2} + \left | \mathbf{b} \right |^{2} \right \rangle}
	\end{equation}
	 and is calculated in the same manner as \cite{Stansby2018c}. $\mathbf{v} = \mathbf{v}_{p} - \mathbf{v}_{p0}$ are the proton velocity fluctuations in the Alfv\'en wave frame, $\mathbf{v}_{p0}$ is chosen to maximise the value of $\left | \sigma_{c} \right |$, and $\mathbf{b} = v_{A} \left ( \mathbf{B} / \left | \mathbf{B} \right | \right )$ is the magnetic field in velocity units. The time averages denoted by $\left \langle \right \rangle$ are taken over all points in non-overlapping 20 minute windows. Uncertainties in $\sigma_{c}$ calculated with a bootstrap method are always less than $\pm 0.1$ and almost always less than $\pm 0.05$.
\end{itemize}

\section{Results}
\begin{figure*}
	\includegraphics[width=2\columnwidth]{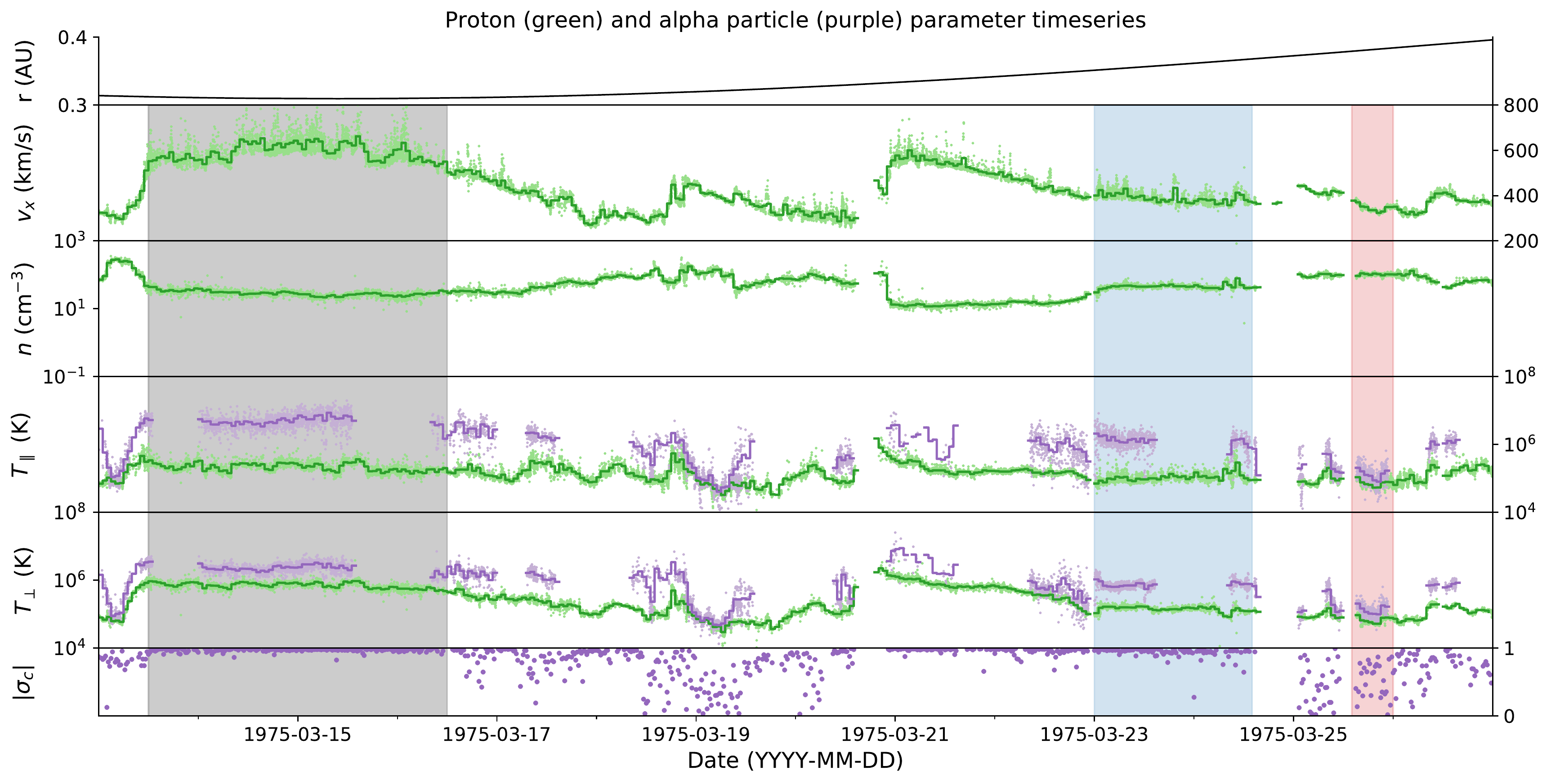}
	\caption{Proton (green) and alpha particle (purple) parameter timeseries from the first perihelion pass of \emph{Helios 1}. From top to bottom: heliocentric distance, radial velocity, number density, parallel temperature, perpendicular temperature, and absolute cross helicity. In the middle 4 panels 40.5 second cadence measurements are shown with light dots, and hourly averaged parameters with dark lines.}
	\label{fig:timeseries}
\end{figure*}

Figure \ref{fig:timeseries} presents an overview of the first perihelion pass of \emph{Helios 1} in early 1975. This shows a typical solar minimum structure in the ecliptic plane, with several fast coronal hole streams interspersed with slower wind speed periods.

The intervals listed in table \ref{tab:intervals} are shown with coloured vertical bands in figure \ref{fig:timeseries}, and are examples of fast (grey), slow Alfv\'enic (blue) and slow non-Alfv\'enic (red) wind. The choice of intervals was primarily motivated by the need to have enough alpha particle measurements to carry out the later statistical analysis. Note that the slow Alfv\'enic stream begins after clear discontinuities in proton number density and temperature, and is clearly distinct from the trailing edge of the preceding high speed stream. Both the fast and slow Alfv\'enic periods have very high Alfv\'enicities, with $\left | \sigma_{c} \right | > 0.9$, whereas the non-Alfv\'enic wind is characterised by a large scatter of $ \left | \sigma_{c} \right |$ values between 0 and 1.

\begin{figure}
	\includegraphics[width=\columnwidth]{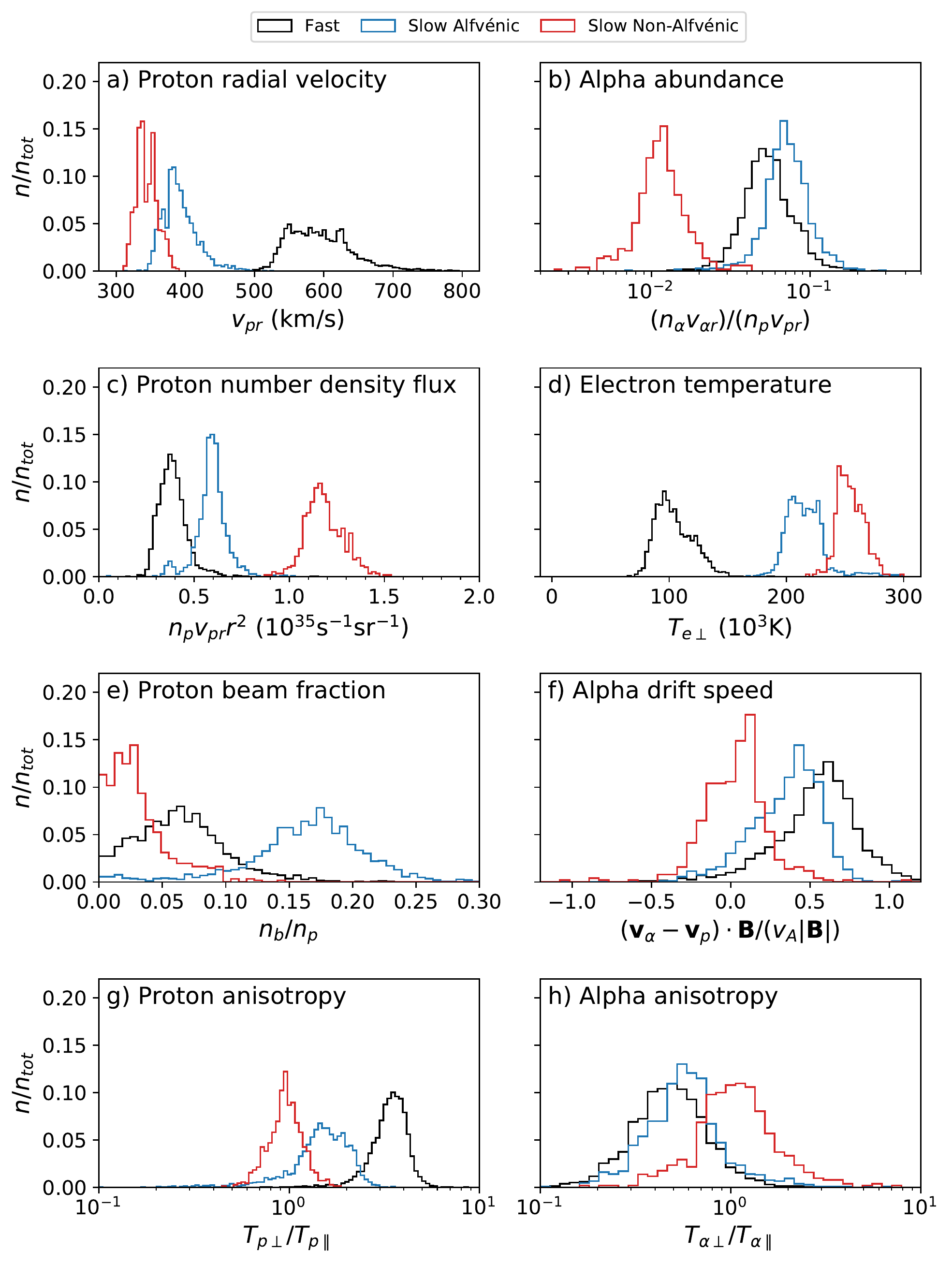}
	\caption{Histograms of various parameters in the three different intervals, normalised to the total number of data points in each interval.}
	\label{fig:stats}
\end{figure}

Figure \ref{fig:stats} shows histograms of various parameters in each type of wind, using the same interval colour coding as figure \ref{fig:timeseries}. Panel a) shows the proton radial velocity, making clear that the slow Alfv\'enic wind has significantly slower speeds ($\sim$ 400 km/s) than the fast wind ($\sim$ 600 km/s).  Panel~b) shows the alpha particle flux, normalised to the proton flux ($v_{\alpha r}n_{\alpha} /v_{pr} n_{p}$). Note that the relative flux is plotted instead of the relative abundance, since the abundance is not conserved if the alpha particle and proton velocities change, whereas the flux is conserved \citep{Hollweg1974, Wang2008}. At these distances the relative fluxes and abundances are similar however, due to relatively similar proton and alpha particle radial speeds. The slow Alfv\'enic wind has similar relative alpha particle fluxes to the fast solar wind ($\sim$ 0.05), whereas the non-Alfv\'enic wind has significantly smaller fluxes, as shown previously (for another interval) by \cite{Ohmi2004}. Panel c) shows the proton number density flux, normalised to radial distance (ie. particles per second per solid angle). The slow Alfv\'enic wind has a flux around twice that of the fast wind, and the non-Alfv\'enic wind has large fluxes around four times the fast wind flux. Panel d) shows the electron perpendicular temperature, which in the slow Alfv\'enic wind (0.2 MK) is twice that of the fast wind (0.1 MK), but slightly cooler than the non-Alfv\'enic wind (0.25 MK). Panel e) shows the proton beam fraction. The non-Alfv\'enic wind has a low proton beam fraction ($<$ 5\%), in contrast to the fast wind ($\sim$ 7\%), whereas the slow Alfv\'enic wind has a significantly higher beam fraction at $\sim$ 18\%. This agrees well with previous estimates of the beam fraction from \emph{Helios}, showing that when a proton beam is present, it has a higher relative density in slower wind \citep{Marsch1982b}. Panel f) shows the alpha particle streaming speed. In agreement with previous studies \citep[e.g.][]{Marsch1982c, Neugebauer1996} alphas stream at a significant fraction of the Alfv\'en speed in the fast solar wind, and the alpha particles also stream at a slightly smaller but still significant fraction in the slow Alfv\'enic wind \citep[as also shown for another interval by][]{Marsch1981}. The non-Alfv\'enic wind has an average proton-alpha drift speed close to zero, which may be due to its relatively high collisionality (due to a high density and low temperature). Finally, panels g) and h) respectively show the proton and alpha particle temperature anisotropy. The non-Alfv\'enic wind is isotropic for both particle populations, again most likely due to a high collisionality. The sense of temperature anisotropy is the same in the fast and slow Alfv\'enic winds, with $T_{\perp} > T_{\parallel}$ for protons but $T_{\perp} < T_{\parallel}$ for alpha particles, however the proton temperature anisotropy is weaker in the slow Alfv\'enic wind compared to the fast wind \citep[see also][]{Stansby2018c}. Note that at least in the fast wind, these features in alpha particle temperatures are not observable at 1 AU, due to the effect of plasma micro-instabilities on the evolution of alpha particle temperatures \citep{Stansby2019a}.

\begin{figure}
	\centering
	\includegraphics[width=0.6\columnwidth]{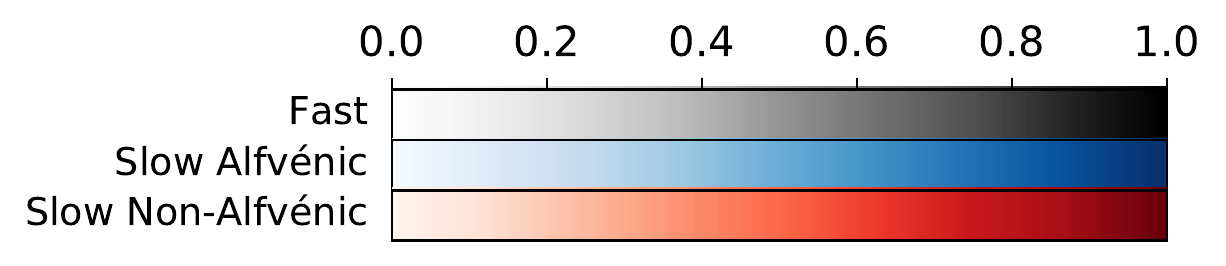}
	\includegraphics[width=\columnwidth]{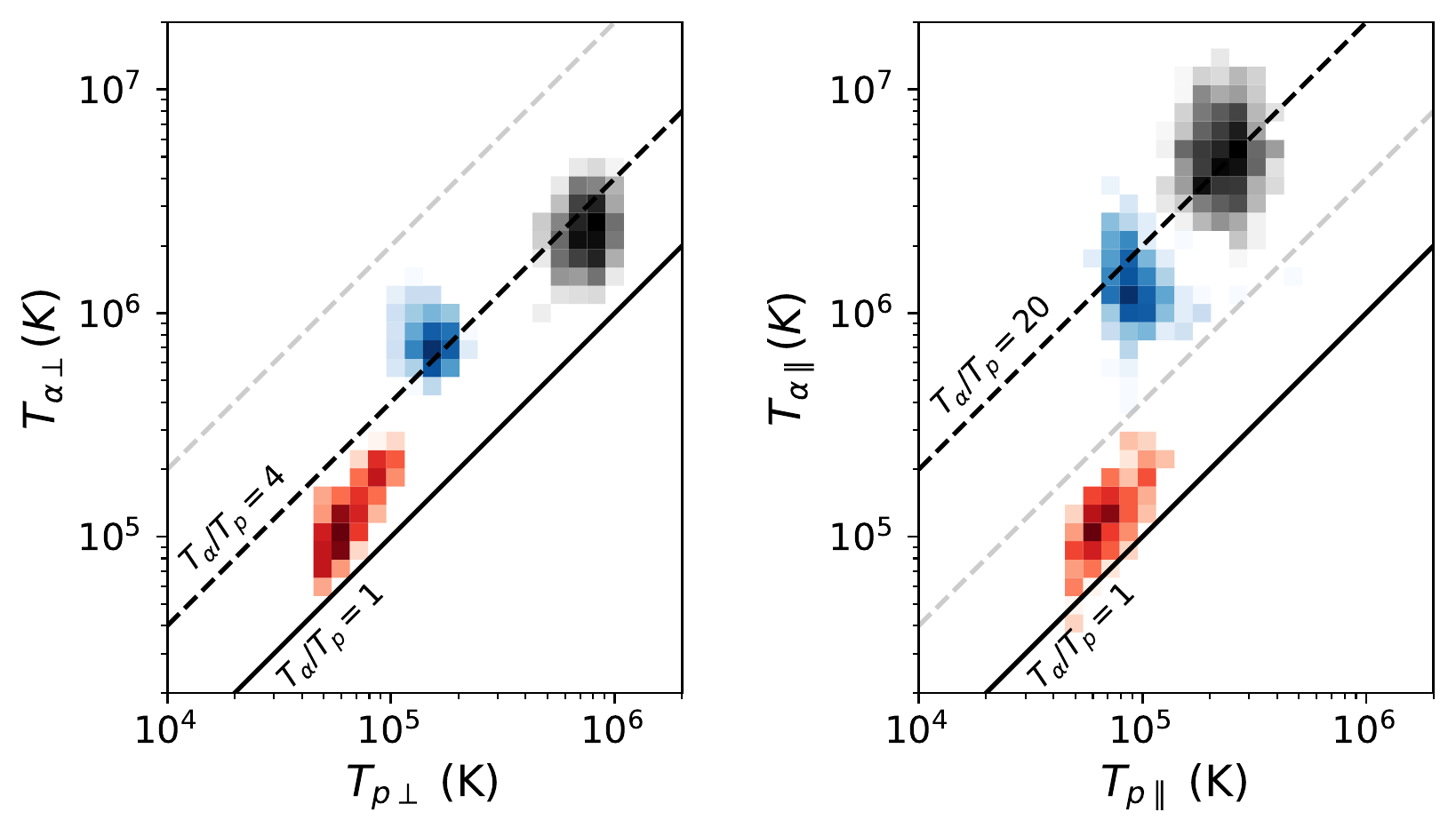}
	\caption{Joint distribution of magnetic field perpendicular (left hand panel) and parallel (right hand panel) proton and alpha particle temperatures in three types of solar wind. Distributions are normalised to the maximum bin value.}
	\label{fig:temps}
\end{figure}

Figure \ref{fig:temps} shows the joint distributions of proton and alpha particle temperatures in each of the three types of wind, with the same colour coding as figure \ref{fig:stats}. The non-Alfv\'enic wind has $T_{\alpha\perp} \approx T_{\alpha \parallel} \approx 10^{5}$ K. The similarity in proton and alpha particle temperatures here may again be due to the non-Alfv\'enic wind's relatively high collisionality. On the other hand the slow Alfv\'enic wind has an order of magnitude larger alpha particle temperatures of $\sim$~$10^{6}$ K. The alpha to proton temperature ratios are similar in the fast and slow Alfv\'enic wind, in both the perpendicular ($T_{\alpha \perp} \approx 4T_{p\perp}$) and parallel ($T_{\alpha \parallel} \approx 20T_{p\parallel}$) directions.

\section{Discussion}
The slow Alfv\'enic solar wind has similar alpha to proton number density fluxes as the fast solar wind (figure \ref{fig:stats}, panel b). Because alpha particles are heavier than protons, they require extra forces above those that accelerate the protons acting on them in order to avoid gravitational settling reducing their abundance to significantly less than the photospheric abundance \citep{Hansteen1994, Basu1995, Asplund2009}. Our results suggest that similar mechanisms driving these enhanced Helium fluxes are active in both the fast and slow Alfv\'enic solar winds, but not in the non-Alfv\'enic slow wind. More specifically, there must be mechanisms other than Coulomb friction, which alone would make the alpha abundance very sensitive to the proton number density flux \citep{Geiss1970a}, in contrast to these observations.

In addition, the slow Alfv\'enic and fast winds both have similar proton-alpha drift speeds (figure \ref{fig:stats}, panel f) and alpha to proton temperature ratios (figure \ref{fig:temps}). Proposed mechanisms to impart these properties include second order effects of Alfv\'en waves \citep{Hollweg1974, Chang1976}, ion cyclotron wave acceleration \citep{Hollweg2002}, or reconnection jets \citep{Feldman1993, Feldman1996}. Although the observations presented here cannot distinguish between these different proposals, they suggest that similar mechanisms are active in both the fast and slow Alfv\'enic wind.

The slow Alfv\'enic wind shows some differences to the fast wind however; it has slower speeds (figure \ref{fig:stats}, panel a), higher mass fluxes (figure \ref{fig:stats}, panel c), higher electron temperatures (figure \ref{fig:stats}, panel d), and lower overall proton and alpha temperatures (figure \ref{fig:temps}). These observations agree with the idea that both the fast and slow Alfv\'enic solar wind were produced on open field lines rooted in coronal holes, but with varying magnetic field geometries in the lower corona. In particular, the amount the magnetic field expands in the low corona (up to 2.5$r_{\odot}$), the footpoint magnetic field strength, and the magnetic field inclination at the solar surface are all thought to be important in shaping the properties of coronal hole wind \citep{Suess1984, Wang1990, Bravo1997, Suzuki2006, Pinto2016, Reville2017}.

The amount of magnetic field expansion sets the location of the critical point where the plasma becomes supersonic, with higher expansion factors resulting in higher Alfv\'enic points \citep{Cranmer2005a, Cranmer2007a}. This in turn alters the effects of heating processes. If heating happens whilst the wind is sub-Alfv\'enic, the mass flux is increased but the speed is not expected to be significantly affected; on the other hand, heating the wind whilst it is supersonic is expected to leave the mass flux unchanged but to increase the flow speed \citep{Leer1980, Wang2009}. This leads to more energy being deposited below the critical point for rapidly diverging magnetic fields, thus resulting in slower speeds but increased mass fluxes \citep{Levine1977, Leer1980, Wang1991}, precisely as observed

The lower ion temperatures in the slow Alfv\'enic wind agree with previous statistical observations of a positive correlation between proton bulk speed and both proton \citep{Elliott2012} and alpha particle \citep{Thieme1989a} temperatures, and also agree with observations at 1 AU of the slow Alfv\'enic wind at solar maximum \citep{DAmicis2018}. Although the mechanisms behind this relation are still not clear, it is reproducible in steady state models of the solar wind \citep{Cranmer2007a, Pinto2017, Usmanov2018}.

A striking feature of the slow Alfv\'enic wind is its large proton beam fraction ($\sim$18\%) compared to the fast wind ($\sim$7\%). Although the proton beam has been previously observed in case studies \citep[e.g.][]{Feldman1973, Feldman1993}, and its drift has been statistically characterised for \emph{Helios} \citep{Marsch1982b, Durovcova2019}, at 1 AU \citep{Alterman2018}, and beyond with \emph{Ulysses} \citep{Matteini2013a}, to our knowledge a large statistical study of proton beam fraction has yet to be carried out\footnote{\cite{Marsch1982b} provides a statistical study of the radial variation of $n_{b}/n_{p}$, but only in cases where the proton beam appears as a distinct peak, therefore not including cases with a weak or non-existent beam.}. Such a study would be helpful in the future to determine what causes the observed large variations in proton beam fraction between different streams.

The higher electron temperatures in the slow Alfv\'enic wind with respect to the fast wind are also in agreement with previous observational in situ results at solar minimum, showing an anti-correlation between proton bulk speed and electron temperature inside 0.3 AU \citep{Marsch1989, Pilipp1990}. A similar anti-correlation has been measured closer to the Sun, with remote sensing measurements showing relatively low electron temperatures above coronal holes where fast solar wind originates, compared to quiet Sun areas where some slow solar wind may originate \citep[see section 12 of][and references therein]{DelZanna2018}. In order to compare our in-situ results to remote sensing measurements however, a study of coronal electron temperature differences between coronal holes of varying sizes would be needed, which to our knowledge does not yet exist.

Taken overall our results provide new lines of evidence that a component of the slow solar wind originates in coronal holes, complementing other recent work that has come to a similar conclusion using different methods. Case studies of heavy ion charge state data measured at 1 AU in the ecliptic plane shows that slow Alfv\'enic wind has a similar composition to fast wind, implying a similar solar origin in coronal holes \citep{DAmicis2018}. In addition, statistical studies of heavy ion fractionation measured at and beyond 1 AU at a large range of solar latitudes show a significant fraction of slow solar wind with similarly low Fe/O ratios as fast wind, again implying that a significant fraction of the slow solar wind comes from over-expanded coronal holes \citep{Stakhiv2015, Stakhiv2016}.

\section{Conclusions}
We have presented a detailed case study comparison between the kinetic properties of protons and alpha particles in the fast, slow Alfv\'enic, and slow non-Alfv\'enic solar wind using data taken by \emph{Helios 1} during its first perihelion passage. The similarity in alpha particle abundance, alpha to proton temperature ratio, and alpha particle drift speed in the slow Alfv\'enic and fast winds adds additional evidence that some slow Alfv\'enic wind originates in coronal holes, similarly to the fast solar wind. The differences in speed, mass flux, and electron temperatures between the slow Alfv\'enic and fast solar wind are explained by different magnetic field geometries in the low corona: the slower wind is released on magnetic field lines that undergo over-expansion that modifies the effects of coronal heating and acceleration processes.

An obvious next step would be performing a magnetic field connectivity analysis \citep[e.g.][]{Neugebauer1998} to determine if the observing spacecraft was really connected to a small coronal hole at the time of measurement. Unfortunately, to our knowledge, there are no magnetic field or extreme ultraviolet images of the Sun available for the interval studied in 1975. However, \emph{Parker Solar Probe} \citep[PSP,][]{Fox2015} is predicted to have been connected to a small coronal hole during its first closest approach to the Sun \citep{Riley2019a}. Having taken in situ measurements of the solar wind down to 0.15 AU, PSP will hopefully allow measurement of the kinetic features of solar wind un-ambiguously emitted from a small coronal hole during solar minimum.
\section*{Acknowledgements}
The authors thank Roberto Bruno for helpful discussions, Marcia Neugebauer for comments and suggestions that significantly contributed to the results and discussion, and the referee for comments that improved discussion of our results. D.~Stansby, T.~S.~Horbury, and D.~Perrone were supported  by STFC grant ST/N000692/1. This work was supported by the Programme National PNST of CNRS/INSU co-funded by CNES.

The authors are grateful to the \emph{Helios} and ACE instrument teams for making the data used in this study publicly available. Data were retrieved using HelioPy v0.8.0 \citep{Stansby2019b} and processed using astropy v3.2.1 \citep{Price-Whelan2018}. Figures were produced using Matplotlib v3.1.1 \citep{Hunter2007}. 

Code to reproduce the figures presented in this paper is available at \url{https://github.com/dstansby/publication-code}.
\bibliographystyle{mnras}
\bibliography{/Users/dstansby/Dropbox/zotero_library}

\end{document}